\documentclass[aps, prb, twocolumn, amssymb, amsmath, superscriptaddress]{revtex4-1}

\usepackage{bm}
\usepackage{times}
\usepackage{graphicx}
\usepackage[colorlinks,citecolor=blue,linkcolor=blue]{hyperref}
\usepackage{array}
\usepackage{float}
\usepackage{colortbl}
\usepackage{caption}
\usepackage{multirow}
\usepackage{tabularx}

\usepackage{amsmath}

\usepackage{gensymb}
\graphicspath{ {./} }

\begin{document}
\title{Novel magnesium borides and their superconductivity}

\author{M. Mahdi Davari Esfahani}
\affiliation{Department of Geosciences, Center for Materials by Design, and Institute for Advanced Computational Science,
State University of New York, Stony Brook, NY 11794-2100, USA}
\author{Qiang Zhu}
\affiliation{Department of Geosciences, Center for Materials by Design, and Institute for Advanced Computational Science,
State University of New York, Stony Brook, NY 11794-2100, USA}
\author{Huafeng Dong}
\affiliation{Department of Geosciences, Center for Materials by Design, and Institute for Advanced Computational Science,
State University of New York, Stony Brook, NY 11794-2100, USA}
\author{Artem R. Oganov}\email{artem.oganov@stonybrook.edu}
\affiliation{Skolkovo Institute of Science and Technology, Skolkovo Innovation Center, 3 Nobel St., Moscow 143026, Russia}
\affiliation{Department of Geosciences, Center for Materials by Design, and Institute for Advanced Computational Science,
State University of New York, Stony Brook, NY 11794-2100, USA}
\affiliation{Moscow Institute of Physics and Technology, 9 Institutskiy Lane,
Dolgoprudny City, Moscow Region 141700, Russia}
\affiliation{International Center for Materials Design, Northwestern Polytechnical University, Xi'an,710072, China}

\author{Shengnan Wang}
\affiliation{Department of Geosciences, Center for Materials by Design, and Institute for
Advanced Computational Science, State University of New York, Stony Brook,
NY 11794-2100, USA}
\author{Maksim S. Rakitin}
\affiliation{Department of Geosciences, Center for Materials by Design, and Institute for
Advanced Computational Science, State University of New York, Stony Brook,
NY 11794-2100, USA}
\affiliation{NSLS-II, Brookhaven National Laboratory, Upton, NY 11973-5000, USA}
\author{Xiang-Feng Zhou}
\affiliation{Department of Geosciences, Center for Materials by Design, and Institute for
Advanced Computational Science, State University of New York, Stony Brook,
NY 11794-2100, USA}
\affiliation{School of Physics and Key Laboratory of Weak-Light Nonlinear Photonics, Nankai University, Tianjin 300071, China}

\date{\today}

\begin{abstract}
With the motivation of searching for new superconductors in the Mg-B system, we performed ab initio evolutionary searches for all the stable compounds in this binary system in the pressure range of 0-200 GPa.
We found previously unknown, yet thermodynamically stable, compositions MgB$_3$ and Mg$_3$B$_{10}$.
 Experimentally known MgB$_2$ is stable in the entire pressure range 0-200 GPa, while MgB$_7$ and MgB$_{12}$ are stable at pressures below 90 GPa and 35 GPa, respectively. We predict a reentrant behavior for MgB$_4$, which becomes unstable against decomposition into MgB$_2$ and MgB$_7$ at 4 GPa and then becomes stable above 61 GPa. We find ubiquity of phases with boron sandwich structures analogous to the AlB$_2$-type structure. 
However, with the exception of MgB$_2$, all other magnesium borides have low electron-phonon coupling constants  $\lambda$  of 0.32 to 0.39 and are predicted to have T$_c$ below 3 K.

\end{abstract}

\keywords{Superconductivity, high pressure, evolutionary algorithm, density functional theory}
\maketitle

\section*{Introduction}
Tremendous efforts have been put to design conventional superconductors with higher and higher critical temperatures 
\cite{Choi2009,Dewhurst2002,Andrzejewski2008,PhysRevB.64.140509,Cava1994a}. 
It is also the main focus of theoretical and experimental studies to determine that how high the superconducting transition temperature T$_c$ can be pushed in binary and ternary boron-compounds. For instance,
theoretical works predicted thermodynamically unstable CaB$_2$ to be superconducting at $\sim$50 K \cite{Choi2009} and hole-doped LiBC to have T$_c$ of 65 K \cite{Dewhurst2002}. Ternary Mo$_2$Re$_3$B with T$_c$= 8.5 K \cite{Andrzejewski2008}, CuB$_{2-x}$C$_x$ (T$_c$ $\sim$ 50 K) \cite{PhysRevB.64.140509} and multiple-phase bulk sample of yttrium-palladium-boron-carbon (T$_c$= 23 K \cite{Cava1994a}) are important boron-based superconductors.

 The unexpected discovery of superconductivity in MgB$_2$  with high T$_c$ = 39 K \cite{nagamatsu2001superconductivity}  has triggered a flurry of publications. In previous studies, superconductivity in MgB$_2$ has been thoroughly investigated
\cite{Welp2003154,Goncharov2003,Choi2002,PhysRevB.65.180517,PhysRevLett.89.107002}. The isotope effect demonstrated the phonon-mediated nature of superconductivity in this compound \cite{Hinks2001}. 
Although doping is usually expressed as a hope to enhance the desired properties, carbon-doped MgB$_2$ (Mg(B$_{0.8}$C$_{0.2}$)$_2$) has a lower T$_c$ = 21.9 K \cite{Ribeiro2003}.
Aluminum, with one more electron than magnesium, was reported to be an unfit candidate for partial substitution for magnesium (Mg$_{1-x}$Al$_x$B$_2$)\cite{Slusky2001}. This shows that increasing electron concentration suppresses superconductivity of magnesium diboride.

Elemental magnesium \cite{Li2010} and boron \cite{Oganov2009} have been shown to exhibit unexpected chemistry under high pressure, raising the motivation of studying their compounds. 
Moreover, materials composed of light atoms could make good conventional  superconductors. 
The Mg-B system was subject to some explorations of superconductivity \cite{PhysRevB.64.092505,PhysRevB.79.054101,317mgb4}. Stability of boron-rich magnesium borides, e.g., MgB$_7$, MgB$_{12}$ and Mg$_{\sim5}$B$_{44}$ has been extensively studied by experiment at ambient pressure \cite{Pediaditakis2010}.  
Borides of similar metals, 
 e.g.,  {\it Ca-B\/} ~\cite{PhysRevB.88.014107} and {\it Li-B\/} ~\cite{ja308492g} and stability of 41 metal borides \cite {VanDerGeest2014184} were studied and new compounds were shown to appear at high pressure.   
 High-pressure phase of MgB$_2$ (KHg$_2$-type structure) was reported to be a poor metal with no superconductivity, highlighting the main role of delocalized bonding of the boron honeycomb layers in the superconducting properties of MgB$_2$ with AlB$_2$-type structure \cite{PhysRevB.79.054101}. 

 To date, there is no comprehensive and systematic theoretical research into the stability and properties of magnesium borides at high pressure. Here, with the knowledge of the important role of magnesium \cite{Ribeiro2003}, crucial existence of honeycomb boron layers \cite{PhysRevB.79.054101} and substantial effect of electron concentration \cite{Slusky2001}, 
we present results of extensive computational searches for stable magnesium borides Mg$_x$B$_y$ and their superconductivity.

\section*{Methods}
Ab initio variable-composition evolutionary method USPEX\cite{Oganov2210932,Oganov01012010,ar1001318,Zhu:sn5112} was applied to the {\it Mg-B\/} system  at 0, 30, 50, 75, 100, 150 and 200 GPa.
This method has the capability of finding possible compositions and the corresponding stable and metastable structures at given pressures, and successfully predicted new phases of MgB$_2$ at high pressure \cite{PhysRevB.79.054101} and new stable phases of different systems like {\it Na-Cl\/}, boron and {\it Na-He \/} \cite{Zhang20122013,Oganov2009,dong2017stable}. High-temperature superconductivity in hydrogen-rich compounds, e.g., Sn-H \cite{esfahani2016superconductivity} and Ge-H \cite{2017arXiv170105600D} were also studied. 
In this method, we created initial generation of structures and compositions randomly with up to 28 atoms in the primitive cell. Subsequent generations were obtained using heredity, transmutation, softmutation, and random symmetric generator \cite{lyakhov2013new}.

Structure relaxations  were carried out using VASP package \cite{PhysRevB.54.11169} in the framework of density functional theory (DFT) adopting PBE-GGA (Perdew-Burke-Ernzerhof generalized gradient approximation) \cite{PhysRevLett.77.3865}. The projector augmented-wave approach (PAW) \cite{PhysRevB.50.17953}  with [He] core for both Mg and B atoms was used to describe the core electrons and their effects on valence orbitals. A plane-wave kinetic energy cutoff of 600 eV and dense Monkhorst-Pack {\it k\/}-points grids with reciprocal space resolution 2$\pi$ $\times$ 0.03 \AA$^{-1}$ was used \cite{PhysRevB.13.5188}.  
 Phonon frequencies and electron-phonon coupling (EPC)  were calculated using   {\sc Quantum ESPRESSO} \cite{QE2009}. 
 PBE-GGA functional is used for this part. A plane-wave basis set with a cutoff of 60 Ry gave a convergence in energy with a precision of 1 meV/atom.
 For electron-phonon coupling, a 6$\times$6$\times$2, 6$\times$6$\times$4 and a 4$\times$4$\times$4 {\it q\/}-point meshes were used for {\it C2/m\/}-MgB$_3$, {\it Amm2\/}-Mg$_3$B$_{10}$ and {\it C2/m\/}-MgB$_4$, respectively. Denser {\it k\/}-point meshes,
12$\times$12$\times$4, 12$\times$12$\times$8 and 8$\times$8$\times$8 were used for the convergence checks of the EPC parameter $\lambda$.
 
\section*{Results}
\subsection*{Search for stable compounds}

Pressure can stabilize new or destabilize the known compounds, and a proper sampling of all promising compositions is needed.
 In Fig.\ref{fig:1}(a)., the enthalpies of formation $\Delta$H$_f$ per atom (with respect to the stable structures of elemental magnesium and boron) are shown in the convex hull form as obtained from all possible compounds.  
Convex hull gives all thermodynamically stable compositions of a multicomponent system, and their enthalpies of formation (per atom). The convex hull (see Fig.\ref{fig:1}(a)) includes all thermodynamically stable states, while unstable ones will always appear above it. 
The distance of an arbitrary compound above the tieline of the convex hull is a measure of its instability, as it shows the decomposition energy of that compound into the nearest stable compounds. The convex hull construction shows that boron-rich compounds are stabilized at high pressure.

Taking our predicted structures/compounds and experimentally known large-cell structures of MgB$_7$, Mg$_{\sim5}$B$_{44}$, MgB$_{12}$ (all three compounds feature B$_{12}$-icosahedra, and for the latter two, we constructed ordered approximants of disordered experimental structures
 - for MgB$_{12}$ containing 388 atoms in the unit cell), we computed the phase diagram of the Mg-B system. At pressures studied here, MgB$_2$, MgB$_3$, Mg$_3$B$_{10}$, MgB$_4$, MgB$_7$ and MgB$_{12}$ have stability fields, making the phase diagram (Fig. \ref{fig:1}(b)) very rich. A recent list of 41 metal borides  presented in Ref \cite{VanDerGeest2014184} at 0 and 30 GPa, clearly demonstrates metal borides often have a variety of stable phases at high pressure.

On increasing pressures metastable compounds, MgB$_3$ and MgB$_6$, get closer to the tieline. Our calculations indicate that at 54 GPa, MgB$_3$ reaches stability and forms the {\it C2/m\/} structure. Unlike MgB$_3$, MgB$_6$ cannot compete with other compounds and remains metastable throughout the entire pressure range  (0 to 200 GPa).

Although MgB$_6$ emerges as a metastable compound from our calculations, we still studied it, keeping in mind recent observation of superconductivity in YB$_6$ \cite{Souma2005}. Moreover, there is experimental evidence for MgB$_6$ as a non-equilibrium phase \cite{Li1497712}.

\subsection*{MgB$_2$}
Some of us studied high-pressure phases of MgB$_2$ using the evolutionary algorithm USPEX \cite{PhysRevB.79.054101}. Our results accord well with that study, as the phase transition happens at 190 GPa. The transition from AlB$_2$-type structure (Fig.\ref{fig:2}) with space group {\it P6/mmm\/} to KHg$_2$-type structure with space group {\it Imma\/}, completely destroys superconductivity. The role of B-B $\pi$-bonded network and charge transfer from Mg to B atoms are explained as having major role in superconducting properties \cite{PhysRevB.79.054101,PhysRevLett.89.107002}.

\subsection*{MgB$_3$}
MgB$_3$, one of the new high-pressure compounds, lies 5 meV/atom above the MgB$_2$-MgB$_7$ tieline at 50 GPa. It becomes stable at 54 GPa and remains stable until 130 GPa in the {\it C2/m\/} phase. Finally {\it Cmcm\/} structure becomes more favorable than all other possible structures up to 200 GPa. AB$_3$ is interestingly a common stoichiometry for metal borides as reported for WB$_3$ \cite{Liang201348}, MnB$_3$ \cite{niu2014variable} and NaB$_3$\cite{VanDerGeest2014184}, however, MgB$_3$ has not been studied yet, neither computationally nor experimentally.

MgB$_3$ stabilizes at high pressure,  while CaB$_3$ \cite{PhysRevB.88.014107} and LiB$_3$ \cite{ja308492g} are not stable even at high pressure. Structural information for the predicted stable MgB$_3$ phases is provided in Table \ref{tab:1--3_struc} and in Fig. \ref{fig:3}.
The metastable layered {\it C2/m\/} phase at pressures below 43 GPa is important, since it has  graphene-like  hexagonal boron pattern, which may be a hint of a potentially superconducting phase.
\subsection*{MgB$_4$}
MgB$_4$ has a remarkable reentrant behavior: this compound is thermodynamically stable in the pressure range 0-4 GPa, then becomes unstable to decomposition into other borides, and then is again thermodynamically stable at pressures \textgreater 61 GPa (Fig. \ref{fig:1}(b)). Below we consider lowest-enthalpy phases corresponding to this composition (see Fig. \ref{fig:5}).

 The {\it Pnma\/} phase of MgB$_4$ is stable at ambient pressure in accord with theoretical \cite{VanDerGeest2014184}  and experimental \cite{JACE:JACE01664} results and remains the most favorable phase up to 31 GPa. Unlike all the other MgB$_4$ phases and most of magnesium borides at different pressure conditions, which are metallic, {\it Pnma\/}-MgB$_4$ is a semiconductor. The predicted phase diagram shows that at 31 GPa the semiconducting state breaks down, and MgB$_4$ transforms into a metallic {\it C2/m\/} (similar to AlB$_2$-type) structure. {\it C2/m\/} has the lowest enthalpy in a narrow pressure range from 31 to 36 GPa. From 36 to 60 GPa, the {\it P$\bar{1}$\/} phase becomes more favorable, and at very high pressures (60 - 200 GPa), high symmetry structure, {\it I4/mmm\/}, becomes stable.

The main feature of {\it I4/mmm\/} and {\it P$\bar{1}$\/} phases is  prisms of boron that hold one or two magnesium atoms.
 Having boron double-layers (in comparison with {\it P6/mmm\/}-MgB$_2$), the {\it C2/m\/} structure can be described as boron sandwich of this composition. Boron sandwiches have graphene-like layer(s) of boron, intercalated by  magnesium atoms. Phonon calculations were performed to check the dynamical stability throughout the Brillouin zone. We did not find any dynamical instability (see Fig. \ref{fig:12ss} and Fig. S9, ESI). Due to high density of states (DOS) at the Fermi level (N(E$_f$)), high-pressure phases of MgB$_4$  can be potential candidates for superconductivity. Electron-phonon coupling (EPC) calculations revealed that among MgB$_4$ phases, only layered {\it C2/m\/}-MgB$_4$ is a superconductor.

MgB$_4$ has analogous stoichiometry to many AB$_4$ systems, e.g. MnB$_4$ \cite{Meng2012,niu2014variable}, CrB$_4$ \cite{PhysRevB.85.144116}, CaB$_4$ \cite{Yahia2008,Liu2010} and so forth. AB$_4$ structures are mostly orthorhombic or tetragonal with 20 atoms per cell. Some of these structures are in BaAl$_4$-type structure with space group {\it I4/mmm\/}\cite{Bruzzone:a06788}. 
 By removing Mg from the prisms, one observes a pattern similar to the $\alpha$-Ga structure of boron \cite{Oganov2009}.

Increasing pressure, we see emergence of a graphene-like boron double-layered phase (MgB$_2$ has a simple hexagonal AlB$_2$-type structure, which is a single-layered phase of this type). The extra layer is located 1.7 {\AA} from the first layer and displaced by 0.8 {\AA} (A$\alpha \beta$ A$\alpha \beta$ ..., A represents Mg and $\alpha$,$ \beta$ are B layers). At 36 GPa, some boron blocks were formed with a pattern of 1 and 2 magnesium atoms per block. Finally, at a higher pressure 60 GPa, the body-centered tetragonal BaAl$_4$-type structure (space group {\it I4/mmm\/}), which is widely adopted among AB$_4$ intermetallic compounds, forms. In {\it I4/mmm\/} structure, magnesium is located in the center of the truncated rectangular prisms made of boron atoms. This structure is similar to {\it Cmcm\/}-MgB$_3$, in which, there are  two magnesium atoms located in each of the truncated rectangular prisms (see Fig. \ref{fig:3}(b). and Fig. \ref{fig:structure_4fig}(a).). 
\subsection*{MgB$_6$}
Although MgB$_6$ is predicted to be stable with respect to decomposition to the elements (Mg and B) \cite{313}, it is not stable against decomposition into elemental boron and MgB$_4$ (see Fig. 1(a) and (b)). 
 Furthermore, in an experimental study at ambient pressure, MgB$_6$ was not found as an individual phase \cite{Moiseev2005}. 

Since intercalated graphite AC$_6$ (A = Mg, Ca, Sr, Ba) \cite{PhysRevB.74.094507,PhysRevLett.106.187002,PhysRevB.74.214513} is superconducting, we searched for the lowest enthalpy MgB$_6$ phases. We observed a hexagonal distorted triple-layered phase, which is the lowest in enthalpy, in the pressure range  15-28 GPa that intrigued us. 
MgB$_6$ forms a recently predicted phase at ambient pressure and remains in this {\it Cmcm\/} structure until 15 GPa \cite{313}, above which a triple-layered structure has lowest enthalpy until 28 GPa (Fig. \ref{fig:6}(c) and (d)). Between 28 GPa and 88 GPa, the {\it R-3m\/} structure becomes more favorable, and eventually, very high pressure imposes a  pattern similar to {\it I4/mmm\/}-MgB$_4$ into {\it P2$_1$/m\/}-MgB$_6$ (see Fig. \ref{fig:6}(a).); This pattern emerges in the pressures greater than 90 GPa in both MgB$_6$ and MgB$_4$. We found the {\it Cmcm\/} structure to be a semiconductor  in agreement with the previous report \cite{313}, whereas the rest of the phases are metallic. The semiconductor-metal transition {\it Cmcm\/} $\rightarrow$ {\it P2$_1$/c\/}  happens at 15 GPa. 

\subsection*{Mg$_3$B$_{10}$}
Mg$_3$B$_{10}$, a boron-rich compound, stable above 55 GPa, has a monoclinic (space group {\it C2/m\/}) phase. Above 83 GPa this phase transforms into the {\it P2/m\/} phase (Fig. \ref{fig:1}(b)). Metastable {\it Amm2\/}-Mg$_3$B$_{10}$, which we predict to have the lowest enthalpy among Mg$_3$B$_{10}$ phases in the pressure range 30-42 GPa, has a layered sandwich structure and is superconducting (Fig. \ref{fig:9}(c) and (d)).


\subsection*{Superconductivity}
Kolmogorov {\it et al.\/}, proposed metal sandwiches consisting of one or more layers of metal and a graphene-like layer of boron i.e., MS-2 and MS-4 with single hexagonal layer of B \cite{PhysRevB.74.224507}. 
 In our study we found, boron sandwiches, new structures with one layer of metal atoms alternating with multiple boron layers. Interestingly, boron sandwich structures are ubiquitous here. 
For example, in MgB$_3$, there is a layered structure with space group {\it C2/m\/} below 43 GPa (see Fig. \ref{fig:1-3-12}.)
 featuring $\alpha A \beta \gamma B$... stacking of B-Mg layers ($A$ and $B$ denotes Mg and $\alpha \beta \gamma$ are B layers).  {\it C2/m\/} structure of MgB$_4$ in 31-36 GPa,
 {\it Amm2\/}  structure of Mg$_3$B$_{10}$ in 30-42 GPa 
and {\it P2$_1$/c\/}  structure of MgB$_6$ in 15-28 GPa 
 (see colored areas in Fig. \ref{fig:1}(b).) also feature boron sandwiches.

Boron sandwiches are layered structures with stackings of [MgB$_2$] and/or [MgB$_4$] blocks (see Fig. \ref{fig:1-3-12}.). For example, Mg$_3$B$_{10}$ can be represented as a [MgB$_2$][MgB$_4$][MgB$_4$]... sequence of layers, and MgB$_3$ can be represented as a [MgB$_2$][MgB$_4$]... .
Superconductivity in MgB$_2$ is mostly related to the boron layers, i.e. B-B $\sigma$ and $\pi$-bonded network, therefore, sandwich borides with hexagonal boron layers might have superconducting properties.
We checked this by electron-phonon coupling calculations.
 Eliashberg spectral function ($\alpha^2F$) calculations lead to the results depicted in Fig. \ref{fig:12ss}. and listed in Table \ref{tab:t_c_boron_sandwiches}. The electron-phonon coupling constants ($\lambda$) for different structures at given pressures, logarithmic averaged phonon frequencies ($\omega_{log}$) and superconducting transition temperatures (T$_c$) are also provided (for more information, see the ESI). Density of states at E$_f$ listed in Table \ref{tab:t_c_boron_sandwiches} shows T$_c$ is higher for boron sandwiches with higher N(E$_f$) per electron. 
One can see from the projected density of states that {\it 2p\/} states of B atoms, located in planar nets, dominate the DOS at the Fermi level (see Fig. \ref{fig:10}. Bands structures and total DOS of other sandwich borides are also provided in the ESI, Fig. S3, S6 and S10)


The critical temperature of superconductivity is estimated from the Allen-Dynes modified McMillan equation\cite{PhysRevB.12.905}:

\begin{equation} 
T_c = \dfrac{	\langle \omega_{log} 	\rangle}{1.2} exp \left({\dfrac{-1.04 (1+ \lambda)}{\lambda - \mu^* ( 1 + 0.62 \lambda)}}\right),
\end{equation} 
 
where $\omega_{log}$ is the logarithmic average phonon frequency and $\mu^{*}$ is the Coulomb pseudopotential, 

\begin{equation} 
\omega_{log} = exp \Big[ \dfrac{2}{\lambda} \int \dfrac{d\omega}{\omega} \alpha^2F(\omega) ln(\omega) \Big]
\end{equation}

 The EPC parameter $\lambda$ is defined as integral involving the spectral function $\alpha^2F$:
\begin{equation} 
\lambda = 2 \int_0^{\infty} \dfrac{\alpha^2 F (\omega)}{\omega} d\omega.
\label{eq-allen}
\end{equation}

 Here we used $\mu^*$ = 0.10 for Coulomb's pseudopotential, as a reasonable value for  most materials \cite{PhysRevLett.78.118,PhysRevLett.104.177005,PhysRevB.84.054543}.  
After MgB$_2$, {\it C2/m\/}-MgB$_4$ has the highest T$_c$ of 2.8 K at zero pressure (it is metastable at 0 GPa). In {\it C2/m\/}-MgB$_4$, high-frequency phonons, mostly by boron atoms,  contributes 80.5\%  to the total EPC parameter, and low-frequency modes are mainly from magnesium vibrations with 19.5\% contribution. Sandwich borides, in general, have high phonon density between 200 to 400 cm$^{-1}$, however, Eliashberg spectral function indicates a poor electron-phonon coupling in this range. 
 Logarithmic average phonon frequencies $\langle\omega_{log}\rangle$  is comparable to that of {\it P6/mmm\/}-MgB$_2$, however, much weaker electron-phonon coupling and lower densities of states at the Fermi level result in very low transition temperatures 0.7-2.8 K. (For more information about phonon band structures, phonon density of states, Eliashberg spectral function and electronic band structures of these phases, see the ESI)

Directly relevant to superconductivity of sandwich borides is the value of DOS at the Fermi level. For example, the DOS is 0.044 states/eV per electron for {\it C2/m\/}-MgB$_4$. This is about half the value of the MgB$_2$ which is 0.084 states/eV per electron. We can see a trend of increasing T$_c$ when we have higher DOS (values are listed in Table  \ref{tab:t_c_boron_sandwiches}). However, other parameters are essential as well. Since logarithmic average phonon frequencies are almost equal, outstanding MgB$_2$ superconductivity can be related to the higher density of states at the E$_f$ mainly from boron {\it p\/}-states and stronger electron-phonon coupling parameter $\lambda$ = 0.73 mainly affected by lower frequency modes. $\lambda$ of other boron sandwiches is about half of the value of MgB$_2$ (see values listed in Table \ref{tab:t_c_boron_sandwiches}), which due to exponential dependence of T$_c$ on $\lambda$, the T$_c$ value of MgB$_2$ is about 10 times higher than other magnesium borides.

\section*{Conclusions}
 Using ab initio evolutionary structure search, we have extended our previous study of MgB$_2$ to other possible Mg-B compounds up to megabar pressures.  
 A remarkable variety of candidate high-pressure ground states has been identified. In this systematic study, under pressures from 0 to 200 GPa, we have found 6 stable compounds, i.e., MgB$_2$, MgB$_3$, MgB$_4$, Mg$_3$B$_{10}$, MgB$_7$ and MgB$_{12}$. Interestingly, MgB$_7$ and MgB$_{12}$, which are reported to be stable at ambient pressure, are not competitive at very high  (above 90 GPa) pressure. In all compounds, at sufficiently high pressures sandwich borides give way to structures with three-dimensional topology.

Most of the predicted stable phases are metallic. No magnesium-rich phases are stable. By decreasing pressure to 0 GPa, the T$_c$ value of {\it C2/m\/}-MgB$_4$ is enhanced and reaches 2.8 K.
The importance of layered structures at the boron-rich end of the Mg-B phase diagram is noteworthy. The valence bands close to and below the E$_f$ are dominated by boron {\it p\/}-states in layered structures. Therefore, EPC calculations are performed and revealed Mg-B sandwich borides are superconducting with T$_c$ of 2.5, 1.0 and 0.7 K for {\it C2/m\/}-MgB$_3$, {\it Amm2\/}-Mg$_3$B$_{10}$ and {\it C2/m\/}-MgB$_4$ at 31, 40 and 33 GPa, respectively.

\section*{Acknowledgements}

We thank DARPA (grant W31P4Q1210008),
the Government of Russian Federation (14.A12.31.0003) and the
Foreign Talents Introduction and Academic Exchange Program
(B08040).
X.F.Z thanks the National Science Foundation of China (grant no. 11174152), the National 973 Program of China (grant no. 2012CB921900), and the Program for New Century Excellent Talents in University (grant no. NCET-12-0278). 

\bibliography{scibib}{}
\bibliographystyle{rsc}
\widetext
\clearpage
\begin{figure}[ht!] 
    \centering
    \caption{
Stability of magnesium borides. (a) Calculated convex hulls at different pressures. $\alpha$-phase, $\gamma$-phase and $\alpha$-Ga-type structures are used for boron \cite{Oganov2009} and for magnesium, hexagonal close-packed (hcp) and body-centered cubic (bcc) structures were used \cite{Li2010}. 
(b) Pressure-composition phase diagram. Solid bars show stable phases, whereas hatched bars indicate metastability. Colored areas illustrate layered structures (boron sandwiches) analogous to AlB$_2$-type structure.   
    }
    \label{fig:1}
\end{figure}

\begin{figure}[ht!]
    \centering
    \caption{Structure of thermodynamically stable MgB$_2$ phase with space group {\it P6/mmm\/}. Projections of layered structure along the (a) [001] and (b) [010] directions.}
    \label{fig:2}
\end{figure}

\begin{figure}[ht!]
    \centering
    \caption{Structures of thermodynamically stable/metastable phases of MgB$_3$ (a) {\it C2/m\/} and (b) {\it Cmcm\/} and (c) projections of layered structure with  space group {\it C2/m\/} along the [001] and (d) [100] directions.}
    \label{fig:3}
\end{figure}

\begin{figure}[ht!]
    \centering
    \caption{
Enthalpy per formula unit relative to the {\it P$\bar{1}$\/} structure as a function of pressure for the best
phases with the MgB$_4$ stoichiometry .
    }
    \label{fig:5}
\end{figure}
 \begin{figure}[h!]
    \centering
    \caption{ Structure of MgB$_4$ phases (a) {\it I4/mmm\/}  (b) {\it P$\bar{1}$\/} and projections of {\it C2/m\/} structure (c) along the [001] and (d)[100] directions.
    }
    \label{fig:structure_4fig}
\end{figure}

\begin{figure}[h!]
    \centering
    \caption{Structure of magnesium hexaboride phases (a) {\it P2$_1$/m\/}  (b) {\it R-3m\/}  and projections of {\it P2$_1$/c\/} structure  (c) along the [010] and (d) [001] directions. Large spheres are Mg atoms and small sphere are Boron atoms.}
    \label{fig:6}
\end{figure}

\begin{figure}[ht!]
    \centering
    \caption{Structure of Mg$_3$B$_{10}$ phases  (a) {\it P2/m\/}, (b)  {\it C2/m\/}  and projections of {\it Amm2\/} structure (c) along the  [001] and (d) [100] directions.}
    \label{fig:9}
\end{figure}

 \begin{figure}[ht!]
    \centering
    \caption{Band structure and partial densities of states for the {\it C2/m\/}-MgB$_4$ structure at ambient pressure.}
    \label{fig:10}
\end{figure}

\begin{figure}[ht!]
    \centering
    \caption{Boron sandwiches in Mg-B compounds. (a) {\it C2/m\/}-MgB$_3$, (b) {\it Amm2\/}-Mg$_3$B$_{10}$ (c) {\it C2/m\/}-MgB$_4$ and (d) {\it P2$_1$/c\/}-MgB$_6$}
      \label{fig:1-3-12}
\end{figure}

 \begin{figure}[ht!]
    \centering
    \caption{Phonon band structure, Eliashberg spectral function $\alpha^2F(\omega)$, the integrated electron-phonon coupling constant $\lambda$($\omega$) and PHDOS of {\it C2/m\/}-MgB$_4$ quenched to atmospheric pressure.}
    \label{fig:12ss}
\end{figure}
\clearpage
\begin{table}
\caption{Computed superconducting T$_c$ of different sandwich borides}
\begin{tabular*}{16.5cm}{@{\extracolsep{\fill}}||c||c|c|c|cc|}

\hline
Structure & MgB$_2$ ({\it P6/mmm\/}) & MgB$_3$ ({\it C2/m\/}) & Mg$_3$B$_{10}$({\it Amm2\/}) & MgB$_4$ ({\it C2/m\/})&\\ \hline
\hline
P (GPa) & 0 &  31 & 40 & 33 & 0 \\
\hline
N(E$_f$) [states/eV per electron]& 0.084 & 0.064 & 0.038 & 0.039 & 0.044 \\

\hline
$\lambda$ & 0.73* & 0.38 & 0.33 & 0.32 & 0.39\\
\hline
$\langle\omega_{log}\rangle$ (K) & 719* & 811 & 843 & 784 & 749\\
\hline
T$_c$ (K) & 27.6* & 2.5 & 1.0 & 0.7 & 2.8\\
\hline 
\end{tabular*}
\label{tab:t_c_boron_sandwiches}
\begin{flushleft}
*T$_c$ of MgB$_2$ is calculated for comparison with other compounds. Note that T$_c$ values in this table are calculated without anharmonicity, using isotropic Eliashberg formalism. T$_c$ for MgB$_2$ is in agreement with Reference \cite{Choi2002}. Higher T$_c$ are expected if anisotropy of the electron-phonon interaction  is included, e.g., account for anisotropy results in overestimation of the T$_c$ of MgB$_2$ to 55 K. On the other hand, anharmonicity of the phonons usually lowers the T$_c$ and in the MgB$_2$ case, it lowers the T$_c$ to 39 K \cite{Choi2002}.  
\end{flushleft}
\end{table}

\clearpage
\begin{table}[h!]
\centering 
\caption{Optimized structures of MgB$_3$}
\begin{tabular*}{8.5cm}{@{\extracolsep{\fill}}cccccc}
\hline
\hline
 Phase         & Lattice  & Atom   & x       &y         & z   \\ 
               &parameters&&& &\\
 \hline
 {\it C2/m\/} [2 f.u.] & a = 2.998 \AA              & Mg(4i)   &  0.7471 &   0.0000 & 0.7982  \\               
  layered      & b = 5.109 \AA              & B$_1$(4g)&  0.0000 &   0.6673 & 0.0000  \\                 
  at 30 GPa    & c = 8.852 \AA              & B$_2$(8i)&  0.5210 &   0.6717 & 0.4055  \\                 
               & $\beta$ = 115.30$^{\circ}$ &          &         &          &         \\
               &                            &          &         &          &         \\
\hline

{\it C2/m\/} [4 f.u.]   & a=7.959\AA  &    Mg$_1$(4i) &   0.4363 & 0.0000 &  0.7000  \\  
 at 50 GPa      & b=2.850\AA  &    Mg$_2$(2c) &   0.0000 & 0.0000 &  0.5000  \\
                & c=10.833\AA &    Mg$_3$(2b) &   0.0000 & 0.5000 &  0.0000  \\
  & $\beta$=116.98 $^{\circ}$ &     B$_1$(4i) &   0.9041 & 0.0000 &  0.1214 \\   
                &             &     B$_2$(4i) &   0.8063 & 0.0000 &  0.2382  \\
                &             &     B$_3$(4i) &   0.7390 & 0.0000 &  0.5339  \\
                &             &     B$_4$(4i) &   0.6943 & 0.0000 &  0.6689  \\
                &             &     B$_5$(4i) &   0.8529 & 0.0000 &  0.8334  \\
                &             &     B$_6$(4i) &   0.7445 & 0.0000 &  0.9430  \\

\hline
{\it Cmcm\/} [2 f.u.]  & a = 2.676 \AA &   Mg(4c)  &  0.0000  &  0.5997 &  0.2500  \\
  at 200 GPa   & b = 11.521\AA &  B$_1$(4c)&  0.0000  &  0.2490 &  0.2500  \\
               & c = 2.668\AA &  B$_2$(4c)&  0.0000  &  0.8302 &  0.2500  \\
               &         & B$_3$(4c)&  0.0000  &  0.9667 &  0.2500  \\
\hline
\end{tabular*}
\label{tab:1--3_struc}
\end{table}
\clearpage

\begin{figure}[ht!]
    \includegraphics[width=0.8\textwidth]{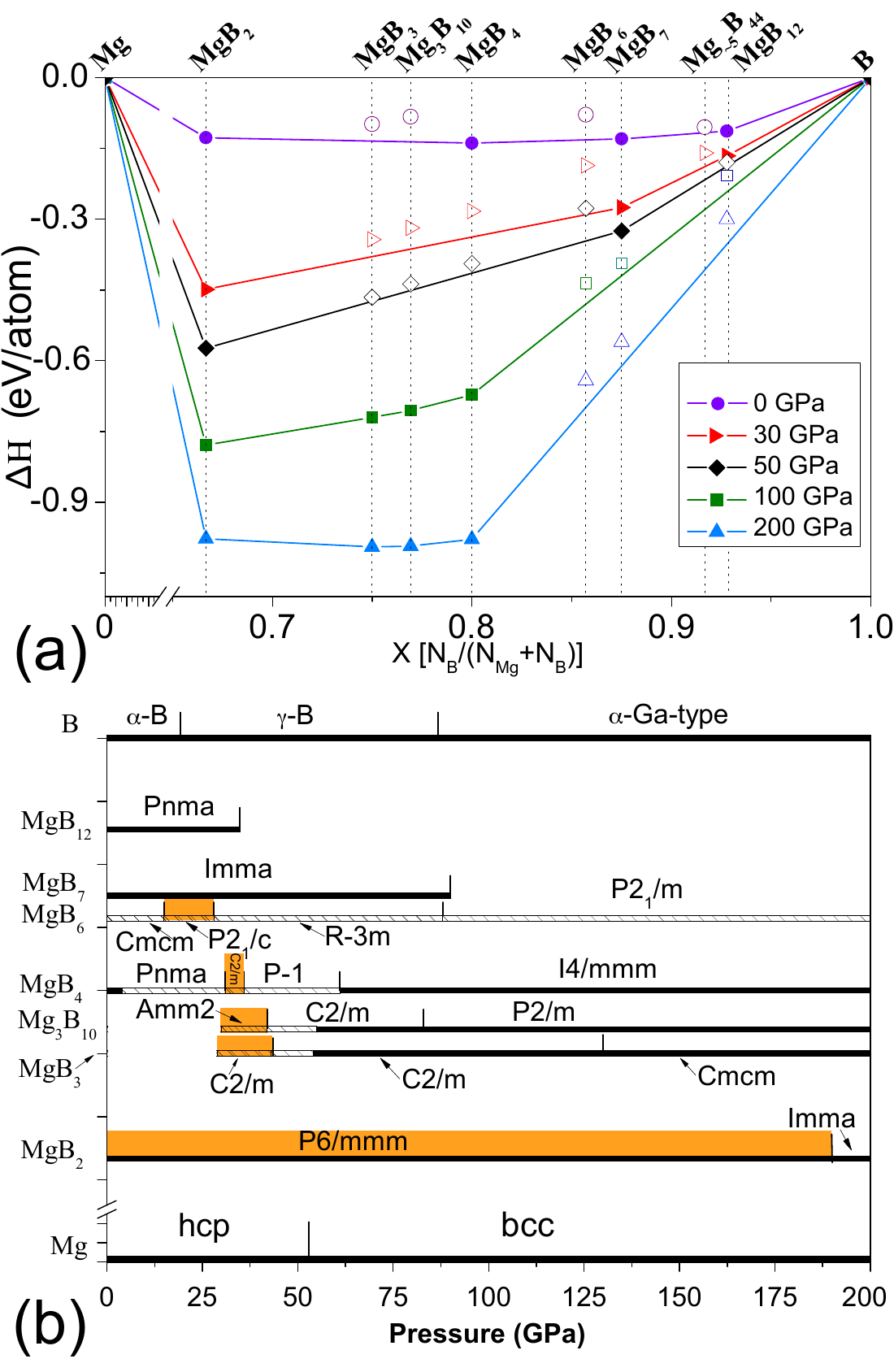}
\end{figure}
\newpage
\begin{figure}[ht!]
    \includegraphics[width=0.8\textwidth]{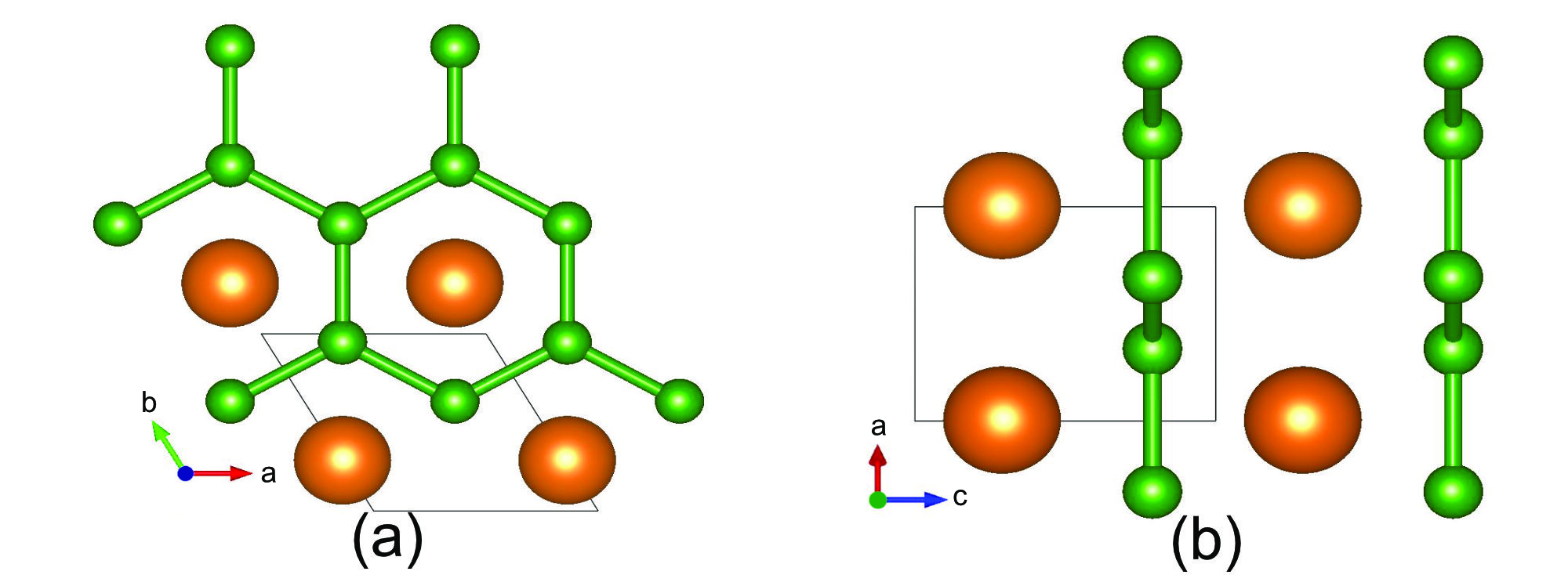}
\end{figure}
 \newpage
\begin{figure}[ht!]
    \includegraphics[width=0.8\textwidth]{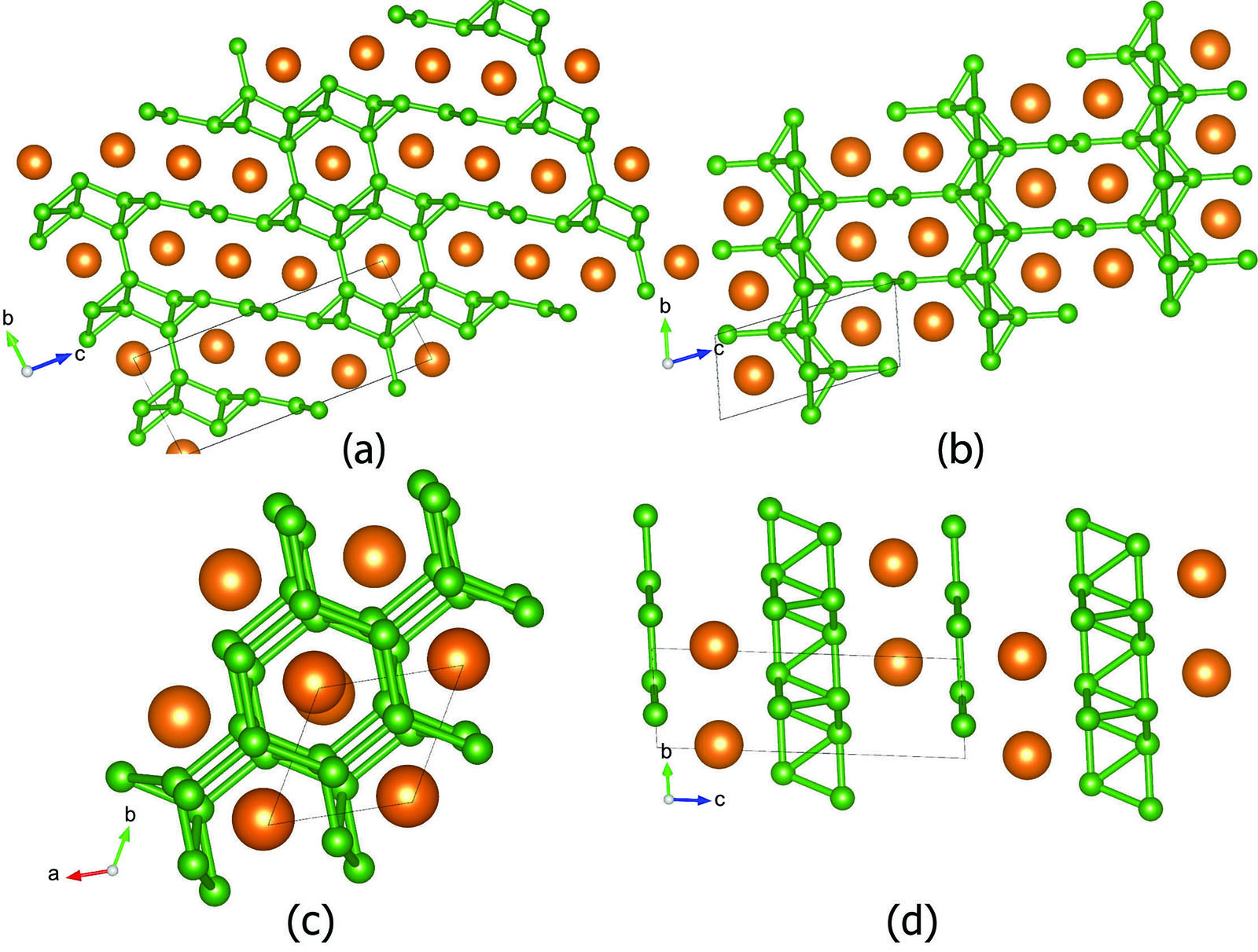}
\end{figure}

 \newpage
\begin{figure}[ht!]
    \centering
    \includegraphics[width=0.8\textwidth]{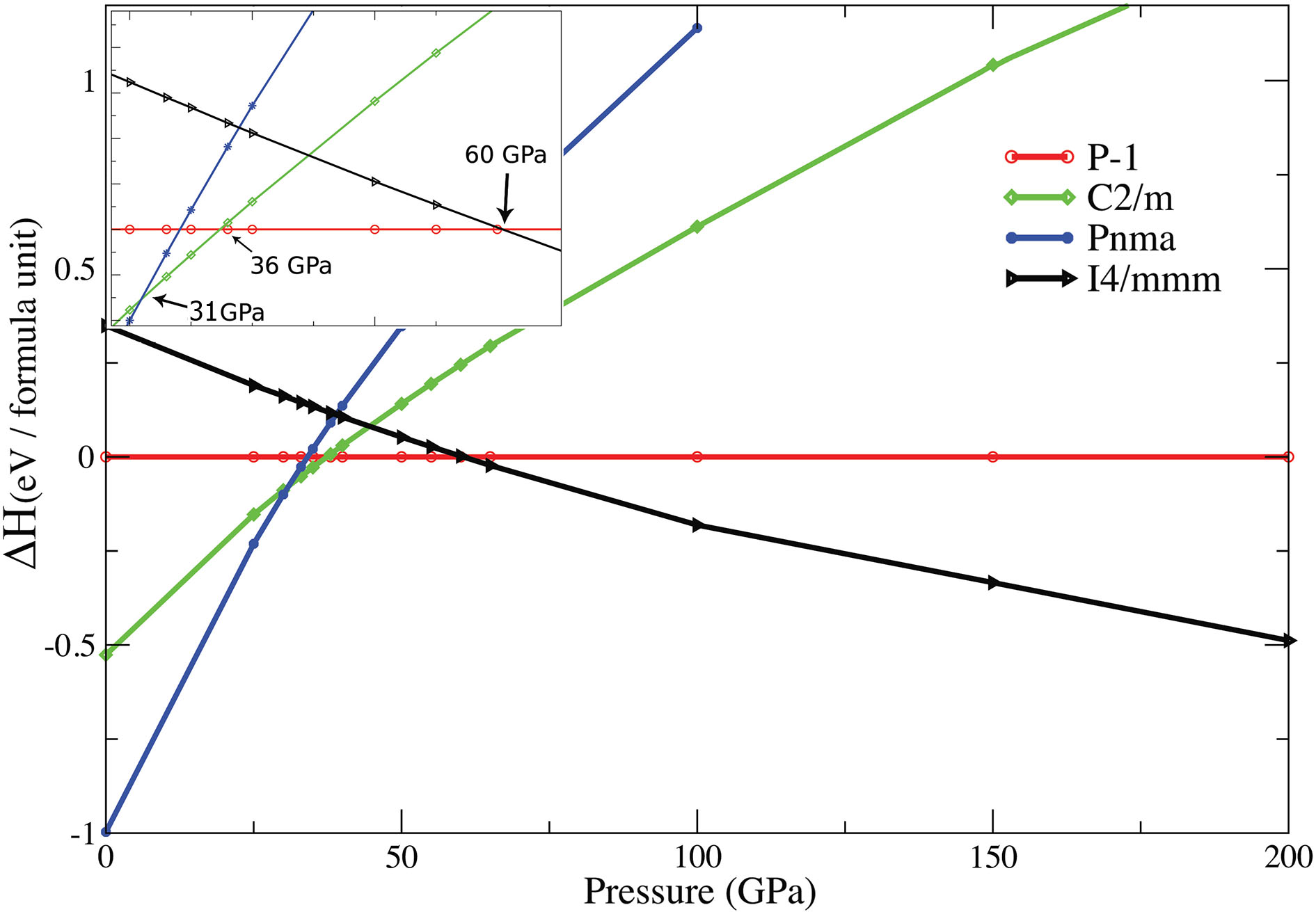}
\end{figure}
 \newpage
 \begin{figure}[h!]
    \centering
    \includegraphics[width=0.8\textwidth]{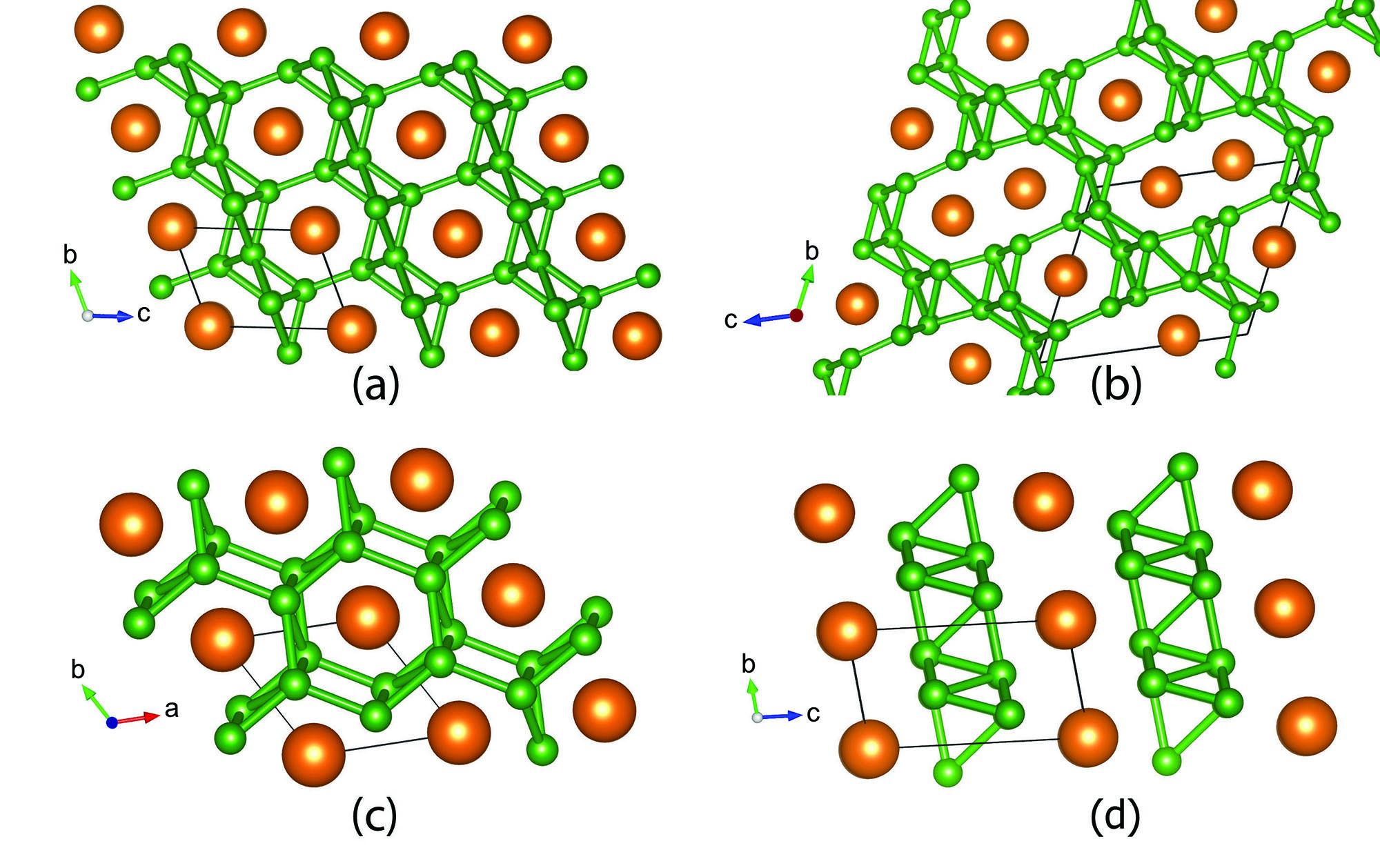}
\end{figure}
 \newpage
\begin{figure}[h!]
    \centering
    \includegraphics[width=0.8\textwidth]{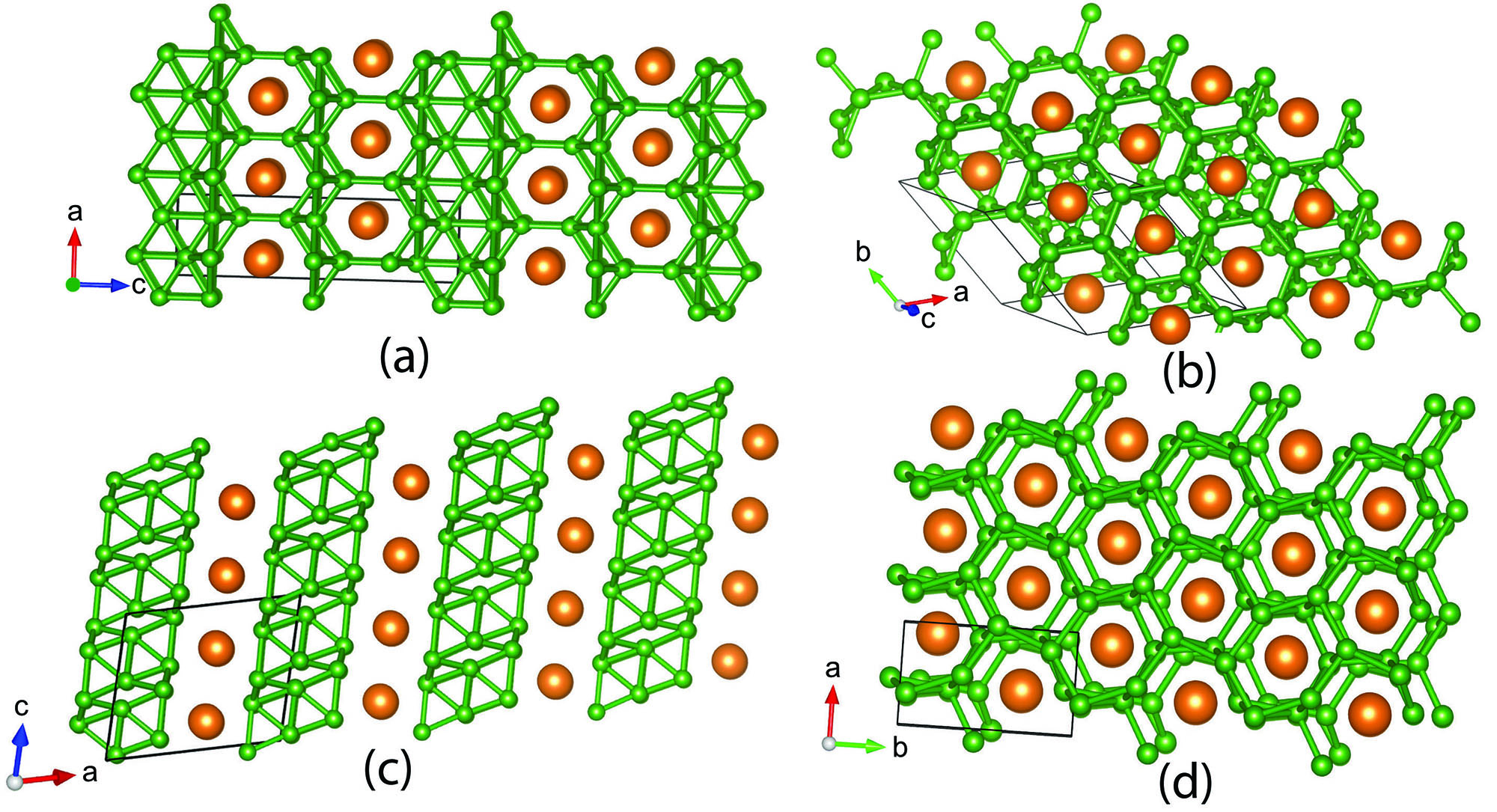}
\end{figure}
 \newpage
\begin{figure}[ht!]
    \centering
    \includegraphics[width=0.8\textwidth]{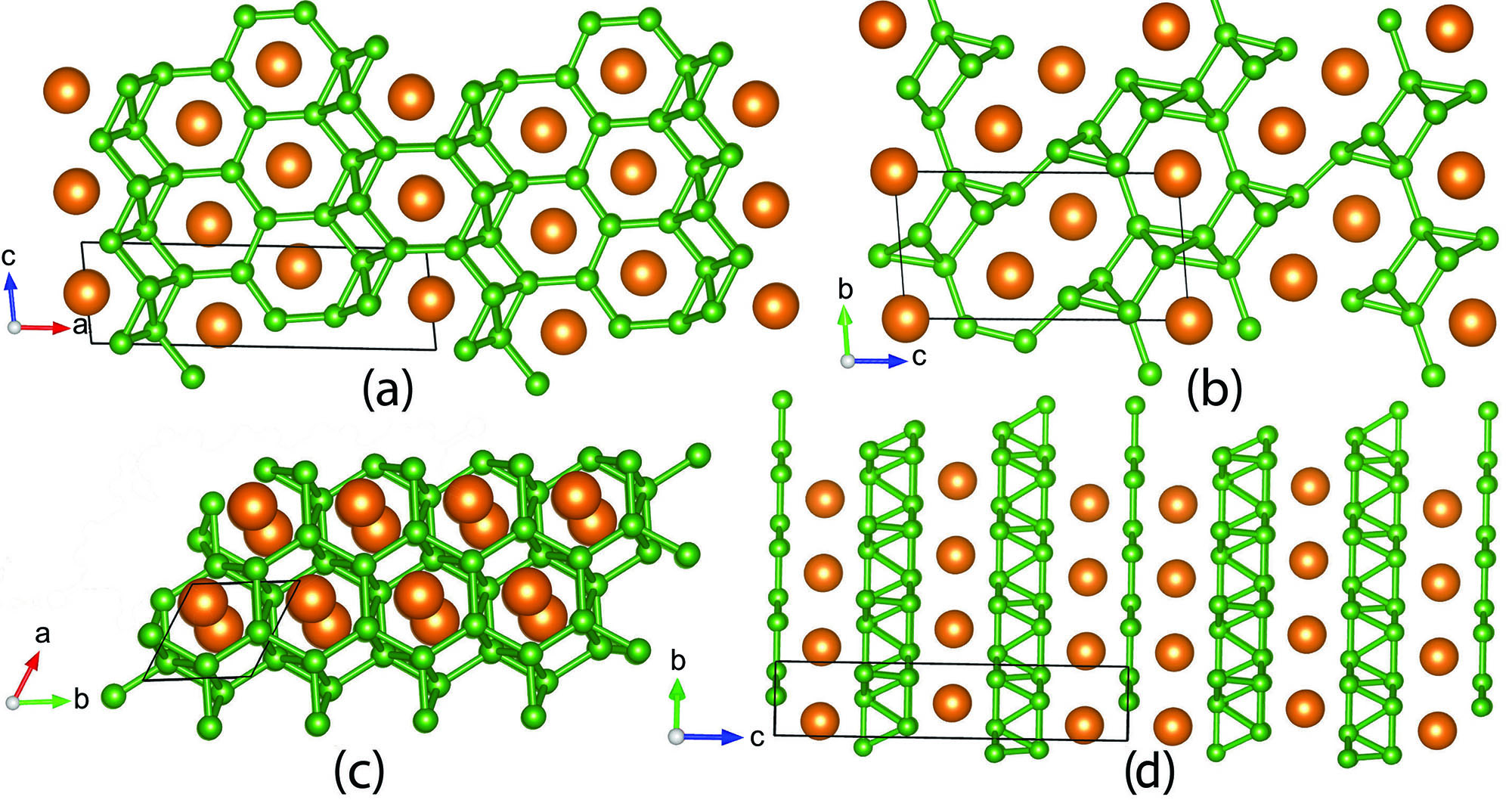}
\end{figure}
 \newpage
 \begin{figure}[ht!]
    \centering
    \includegraphics[width=0.8\textwidth]{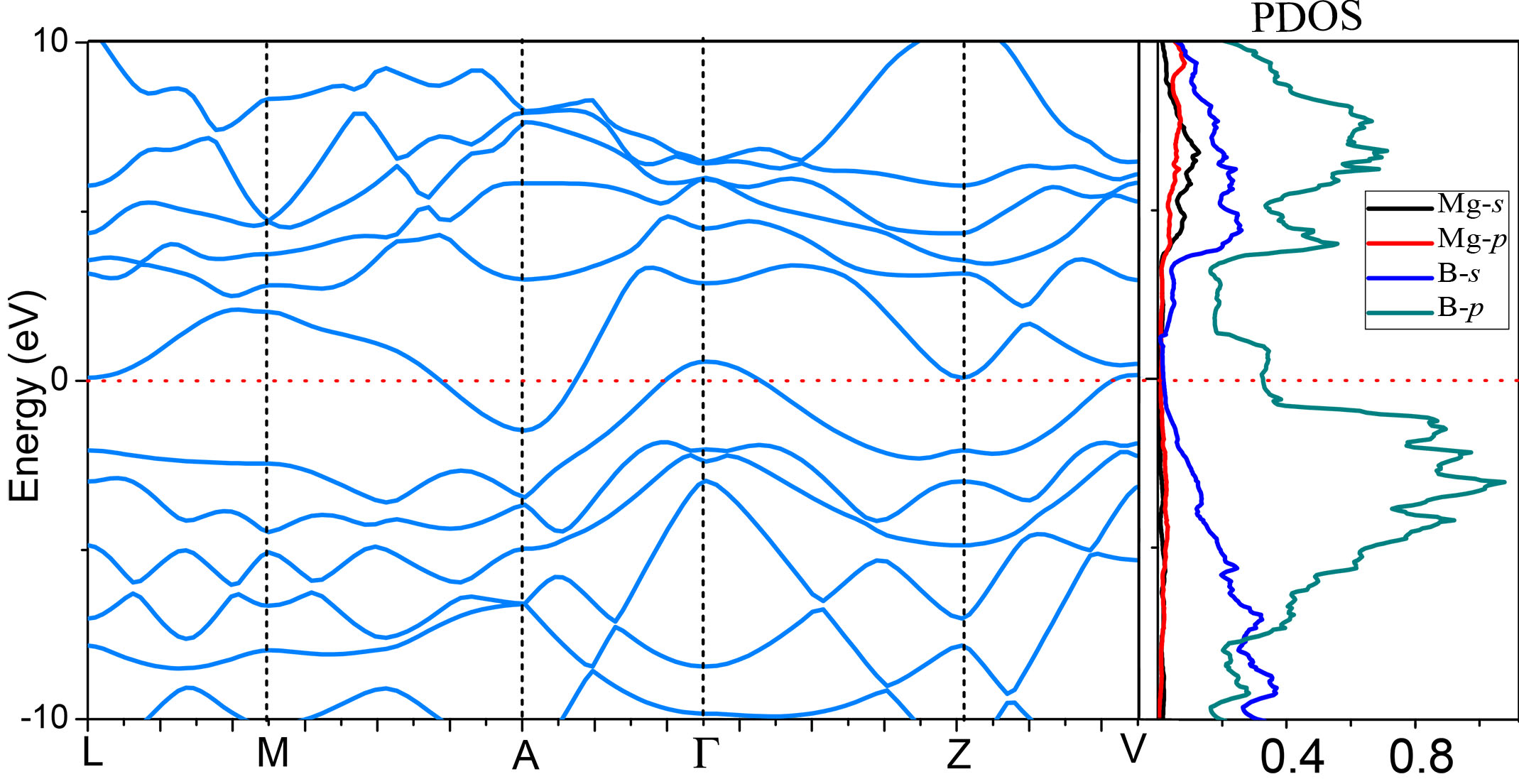}
\end{figure}
 \newpage

\begin{figure}[ht!]
    \centering
    \includegraphics[width=0.8\textwidth]{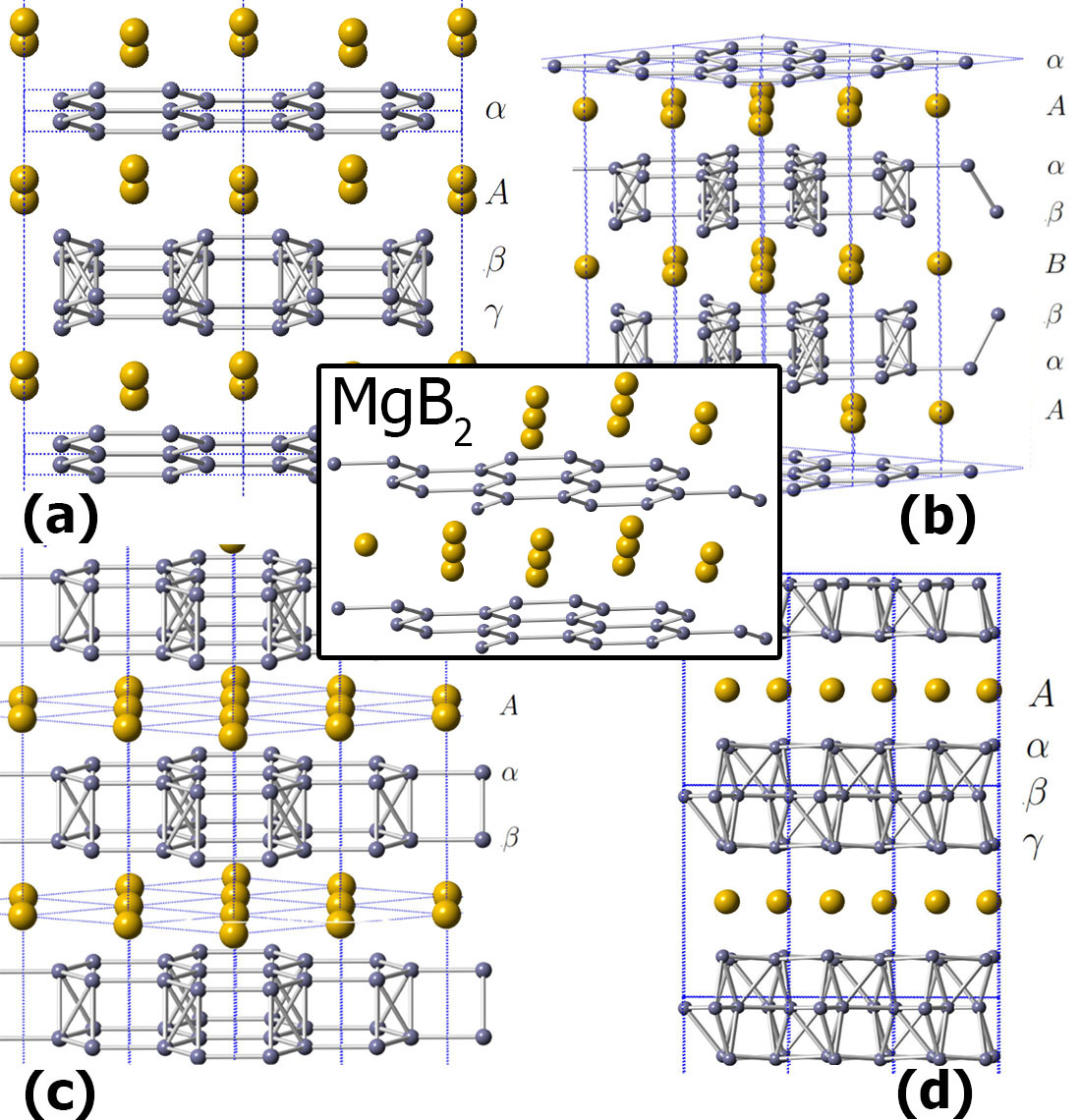}
\end{figure}

 \newpage
 \begin{figure}[ht!]
    \centering
    \includegraphics[width=0.8\textwidth]{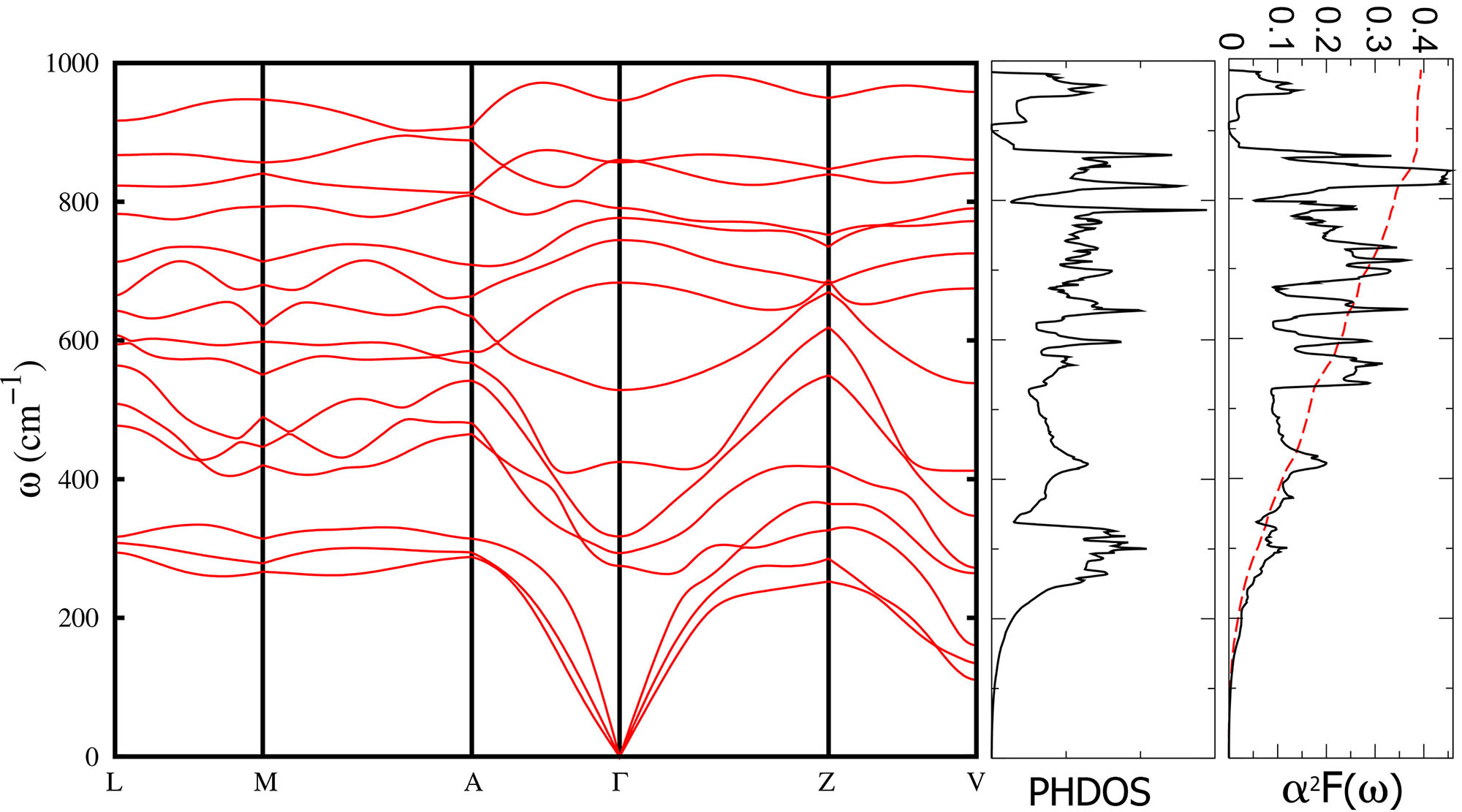}
\end{figure}

\newpage

\widetext

\end{document}